\documentclass{ws-ijmpa-preprint-v6}     
\usepackage[super,compress]{cite}
\usepackage{graphicx}

\usepackage{amssymb,epsf,latexsym,amssymb,amsmath,mathrsfs}
\usepackage[english]{babel}

\newcommand  {\journalref}{Mod. Phys. Lett. A \textbf{29}, 1430018 (2014)}

\newcommand{\beq}{\begin{equation}}
\newcommand{\eeq}{\end{equation}}
\newcommand{\beqa}{\begin{eqnarray}}
\newcommand{\eeqa}{\end{eqnarray}}
\newcommand{\bsubeqs}{\begin{subequations}}
\newcommand{\esubeqs}{\end{subequations}}

\newcommand{\fracnew}[2]
           {\protect\frac{{#1}_{\protect\vphantom{!_A}}}
           {{#2}^{\protect\vphantom{A}}}}

\makeatletter 
\@addtoreset{equation}{section}
\renewcommand{\theequation}{\thesection.\arabic{equation}}
\makeatother

\begin{document}
\markboth{F.R. Klinkhamer}
{A new type of nonsingular black-hole solution in general relativity}

%
\catchline{}{}{}{}{}
%

\title{A NEW TYPE OF NONSINGULAR BLACK-HOLE SOLUTION\\IN GENERAL RELATIVITY}

\author{F.R. KLINKHAMER}

\address{Institute for Theoretical Physics, Karlsruhe Institute of
Technology (KIT),\\ 76128 Karlsruhe, Germany\\
frans.klinkhamer@kit.edu}

\maketitle


\begin{abstract}
\\
Certain exact solutions of the Einstein field equations over
nonsimply-connected manifolds are reviewed. These solutions
are spherically symmetric and have no curvature singularity.
They provide a regularization of the standard Schwarzschild
solution with a curvature singularity at the center.
Spherically symmetric collapse of matter in $\mathbb{R}^4$ may
result in these nonsingular black-hole solutions,
if quantum-gravity effects allow for topology change
near the center or if nontrivial topology is already
present as a remnant from a quantum spacetime foam.
\\
\end{abstract}

\ccode{PACS numbers: 04.20.Cv; 02.40.Pc; 04.20.Jb.}


\section{Introduction}
\label{sec:Introduction}

The main topic of this Brief Review concerns a nonsingular
black-hole solution of general relativity,
which is closely related to
(but not identical with) the standard  Schwarzschild
solution\cite{Schwarzschild1916,HawkingEllis1973,MTW1974}.
It was arrived at by a detour which is rather interesting
by itself.

That earlier investigation started from the following simple question:
\emph{how smooth is space and what quantitative bounds can be set?}
In order to get a first partial answer to this question,
certain Swiss-cheese-type spacetime models were considered,
for which the photon propagation can be calculated in the long-wavelength
limit\cite{Bethe1944,BernadotteKlinkhamer2007}.
Specifically, these spacetime models  have
randomly-positioned identical static defects.

The simplest type of defect is obtained as follows:
take 3-dimensional Euclidean space,
remove the interior of a ball ($r<b$),
and identify antipodal points on the boundary ($r=b$).
The corresponding Swiss-cheese-type model then has two parameters:
the defect size $b$ and the average distance $d$ between neighbouring defects.
The photon propagation over this spacetime model
is described by the isotropic modified Maxwell theory
with a single Lorentz-violating  parameter\cite{BernadotteKlinkhamer2007}
\beq\label{eq:LV-parameter}
\widetilde{\kappa}_\text{tr}
= \pi\;b^3/d^3\,.
\eeq

The isotropic modified Maxwell theory with
parameter $\widetilde{\kappa}_\text{tr}>0$
[phase velocity of light equal to
$\sqrt{(1-\widetilde{\kappa}_\text{tr})/(1+\widetilde{\kappa}_\text{tr})}\,c$
$<$ $c$\,] and the standard Dirac theory of charged particles
[proton limiting velocity equal to $c$\,]
give rise to vacuum-Cherenkov radiation.
From the observed absence of this nonstandard decay process
in the Auger data on ultrahigh-energy cosmic rays, the following
two-$\sigma$ upper bound
has been obtained:\cite{KlinkhamerSchreck2008}
\beq\label{eq:SMEbounds-isotropic-upper}
\widetilde{\kappa}_\text{tr} \;<\; 6 \times 10^{-20}\,.
\eeq
Two remarks are in order:
\begin{romanlist}[(ii)]
\item
Bound (\ref{eq:SMEbounds-isotropic-upper}),
with the particular identification \eqref{eq:LV-parameter},
also holds for $b$ and $d$ close to $L_\text{Planck}$
$\equiv$ $(\hbar\,G_N/c^3)^{1/2}$
$\sim$ $10^{-35}\;\text{m}$.
\vspace*{2mm}\item
The extremely small number on the right-hand side of
\eqref{eq:SMEbounds-isotropic-upper} then suggest that,
whatever the ultimate theory of quantum gravity may be,
this quantum theory must leave practically no defects/wrinkles/ripples
on the emerging classical flat spacetime.
\end{romanlist}

All this is quite intriguing, but the particular defect naively embedded
in standard Minkowski spacetime does \emph{not} satisfy the
vacuum Einstein field equations 
(there are delta-function-type curvature singularities\cite{Schwarz2010} at $r=b$) 
and the same holds for the
corresponding Swiss-cheese-type spacetime models.
The task, then, is to find a proper defect solution.
It turns out that the construction of this nonsingular defect
solution\cite{KlinkhamerRahmede2013,Klinkhamer2014,KlinkhamerSorba2014}  
produces an interesting spin-off: a black-hole solution without curvature
singularity\cite{Klinkhamer2013-MPLA,Klinkhamer2013-APPB}.

The outline of the present review  is as follows.
In Sec~\ref{sec:Topology}, the relevant topology is discussed.
In Sec~\ref{sec:Nonsingular-Defect-Solution},
a nonsingular defect solution is presented, which has no
curvature singularity at the center.
In Sec~\ref{sec:Nonsingular-BH-Solution},
the corresponding black-hole solution is discussed,
which has an event horizon but still
no curvature singularity at the center.
In Sec~\ref{sec:Nonsingular-BH-Solution-with-Electric-Charge},
a related black-hole solution is presented
with a small but nonzero electric charge.
In Sec~\ref{sec:Discussion}, questions related to physics are discussed.
There are also four appendices dealing with technical issues.
Two of these appendices contain some important results:
the simplest possible description of the nonsingular black-hole solution
is given in \ref{app:Nonsingular-solution-in-Different-coordinates}
and an essential physics point is discussed in
\ref{app:Weakened-Elementary-Flatness-Condition}.
A shorter version of this review can be found
in Ref.~\citen{Klinkhamer2013-KSM}.

Let us emphasize, right from the start,
that the solutions discussed in this Brief Review
are solutions of standard general relativity, no more, no less.
The only ``new'' ingredient is the nontrivial topology,
whereas the matter sector is kept entirely standard
(e.g., the Maxwell theory of electromagnetism).
In this respect, our nonsingular black-hole solutions
differ from regular solutions obtained
with a particular nonlinear extension of
electromagnetism\cite{AyonBeatoGarcia1998,Garcia-etal2013}.
(An extensive list of references on regular black holes
can be found in, e.g., Ref.~\citen{Garcia-etal2013}.)

\section{Topology}
\label{sec:Topology}

The four-dimensional spacetime manifold considered is
\bsubeqs\label{eq:M4-M3}
\beq\label{eq:M4}
\widetilde{\mathcal{M}}_4 = \mathbb{R} \times \widetilde{\mathcal{M}}_3\,,
\eeq
where $\widetilde{\mathcal{M}}_3$ is a noncompact,
orientable, nonsimply-connected manifold without boundary.
Up to a point, $\widetilde{\mathcal{M}}_3$ is homeomorphic
to the 3-dimensional real-projective space,
\beq\label{eq:M3-topology}
\widetilde{\mathcal{M}}_3 \simeq
\mathbb{R}P^3 - \{\text{point}\}\,.
\eeq
\esubeqs
Recall that the 3-dimensional real projective space
is topologically equivalent to a 3-sphere with antipodal
points identified.
Here, and in the following, the notation is as follows:
$\widetilde{\mathcal{M}}$ with tilde stands for
a nonsimply-connected manifold [having a nontrivial first homotopy group,
$\pi_1(\widetilde{\mathcal{M}})\ne 0$] and
$\mathcal{M}$ without tilde stands
for a simply-connected manifold
[having a trivial first homotopy group, $\pi_1(\mathcal{M})=0$].

\begin{figure}[t] 
\begin{center}
\includegraphics[width=0.5\textwidth]{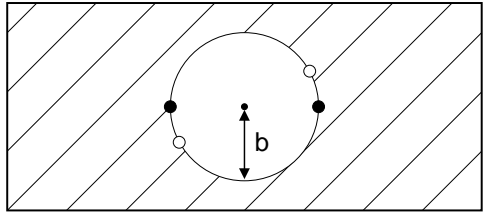}
\end{center}
\caption{Three-space $\widetilde{\mathcal{M}}_3$
obtained by surgery on the 3-dimensional Euclidean space $E_3$:
the interior of the ball with radius $b$ is removed
and antipodal points on the boundary of the ball are
identified (as indicated by open and filled circles).
In $E_3$, the ``long distance'' between antipodal points
is given by $\pi\, b$ and the ``short distance'' by $2\, b$.}
\label{fig:defect}
\end{figure}

For the explicit construction of $\widetilde{\mathcal{M}}_3$, we perform local
surgery on  the 3-dimensional Euclidean space
$E_3=\big(\mathbb{R}^3,\, \delta_{mn}\big)$.
We use the standard Cartesian and spherical
coordinates on $\mathbb{R}^3$,
\bsubeqs\label{eq:Cartesian-spherical-coord}
\beqa
\vec{x}= (x^1,\,  x^2,\, x^3)
= (r \sin\theta  \cos\phi,\,r \sin\theta  \sin\phi,\, r \cos\theta )\,,
\eeqa
with ranges
\beqa
&&
x^m \in (-\infty,\,+\infty)\,,\\[2mm]
&&
r \in [0,\,\infty)\,,\quad
\theta \in [0,\,\pi]\,,\quad
\phi \in [0,\,2\pi)\,.
\eeqa
\esubeqs
Now, $\widetilde{\mathcal{M}}_3$ is obtained from $\mathbb{R}^3$ by
removal of the interior of the ball $B_b$ with radius $b$ and
identification of antipodal points on the boundary $S_b \equiv \partial B_b$.
With point reflection denoted $P(\vec{x})=-\vec{x}$,
the 3-space $\widetilde{\mathcal{M}}_3$ is given by
\beq\label{eq:M3-definition}
\widetilde{\mathcal{M}}_3 =
\Big\{  \vec{x}\in \mathbb{R}^3\,:\; \Big(|\vec{x}| \geq b >0\Big)
\wedge
\Big(P(\vec{x})\cong \vec{x} \;\;\text{for}\;\;  |\vec{x}|=b\Big)
\Big\}\,,
\eeq
where $\cong$ stands for point-wise identification (Fig.~\ref{fig:defect}).

%
%

The next step is to identify \emph{appropriate coordinates} and
a careful analysis has been given in Ref.~\citen{Schwarz2010}.
The standard coordinates of Euclidean 3-space are unsatisfactory,
because a single point of  $\widetilde{\mathcal{M}}_3$
may have different coordinates.
For example, $(x^1,\,  x^2,\, x^3)$ $=$ $(b,\,0,\,0)$
and $(x^1,\,  x^2,\, x^3)$ $=$ $(-b,\,0,\,0)$
correspond to the same point of  $\widetilde{\mathcal{M}}_3$.

\begin{figure}[b] 
\begin{center}    
\includegraphics[width=0.6\textwidth]{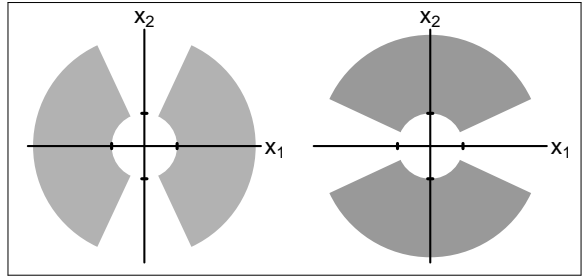}
\end{center}
\caption{Slice $x_3 = 0$ of the manifold $\widetilde{\mathcal{M}}_3$ 
with the domains of the chart-1 coordinates (left) 
and the chart-2 coordinates (right). The tick marks on the
$x_1$ and $x_2$ axes correspond to the values $\pm b$;
see Fig.~\ref{fig:defect}. The 3-dimensional
domains are obtained by revolution
around the $x_1$-axis (left) or the $x_2$-axis (right).
The domain of the chart-3 coordinates is defined similarly.}
\label{fig:defect-charts}
\vspace{0cm}
\end{figure}

A relatively simple choice of coordinates
for $\widetilde{\mathcal{M}}_3$
uses three overlapping charts, each one centered
around one of the three Cartesian coordinate axes.
The manifold $\widetilde{\mathcal{M}}_{3}$
is now covered by three coordinates charts,
\beq\label{eq:XnYnZn}
(X_n,\,  Y_n,\, Z_n)\,, 
\eeq
for $n=1,\,2,\,3$.  These coordinates have the following ranges:
\bsubeqs\label{eq:XnYnZn-ranges} 
\beqa\label{eq:X1Y1Z1-ranges}
X_{1} \in (-\infty,\,\infty) \,,\quad
Y_{1} \in (0,\,\pi)\,,\quad
Z_{1} \in (0,\,\pi)\,,
\eeqa
\beqa\label{eq:X2Y2Z2-ranges}
X_{2} \in (0,\,\pi)\,,\quad
Y_{2} \in (-\infty,\,\infty)\,,\quad
Z_{2} \in (0,\,\pi)\,,
\eeqa
\beqa\label{eq:X3Y3Z3-ranges}
X_{3} \in  (0,\,\pi)\,,\quad
Y_{3} \in (0,\,\pi)\,,\quad
Z_{3} \in  (-\infty,\,\infty)\,.
\eeqa
\esubeqs
In each chart, there is one radial-type coordinate with infinite range,
one polar-type angular coordinate of finite range, and one
azimuthal-type angular coordinate of finite range.
The charts overlap: see Fig.~\ref{fig:defect-charts} for a sketch
and \ref{app:Coordinate-Charts} for further details.

In the following, we will explicitly work with only one coordinate chart,
which we take to be (\ref{eq:X2Y2Z2-ranges}).
Moreover, we will drop the suffix `2' on these coordinates:
\beq\label{eq:TXYZ-def}
(T,\, X,\,  Y,\, Z) \equiv(T,\, X_2,\,  Y_2,\, Z_2)\,,
\eeq
where the time coordinate $T$ has been added in order
to describe part of the spacetime manifold $\widetilde{\mathcal{M}}_4$.

\begin{figure}[h] 
\begin{center}
\vspace*{-.50mm}
\includegraphics[width=0.5\textwidth]{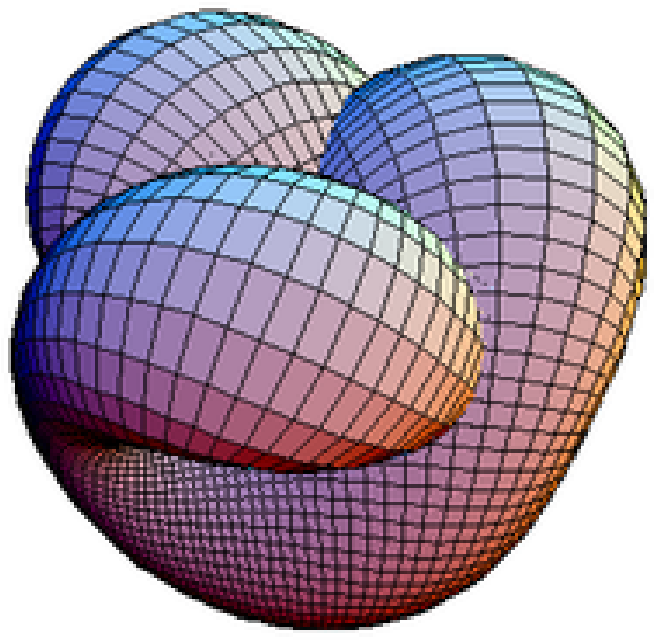}
\end{center}
\caption{Boy's surface immersed in $\mathbb{R}^3$.
Boy's surface is homeomorphic to the real projective plane, $\mathbb{R}P^2$.
Figure obtained by use of the Bryant--Kusner parametrization
of $\mathbb{R}P^2$ and the computer program
\textsc{MATHEMATICA}\cite{Weisstein}.}
\label{fig:Boy's-surface}
\vspace*{.5mm}
\begin{center}
\includegraphics[width=0.35\textwidth]{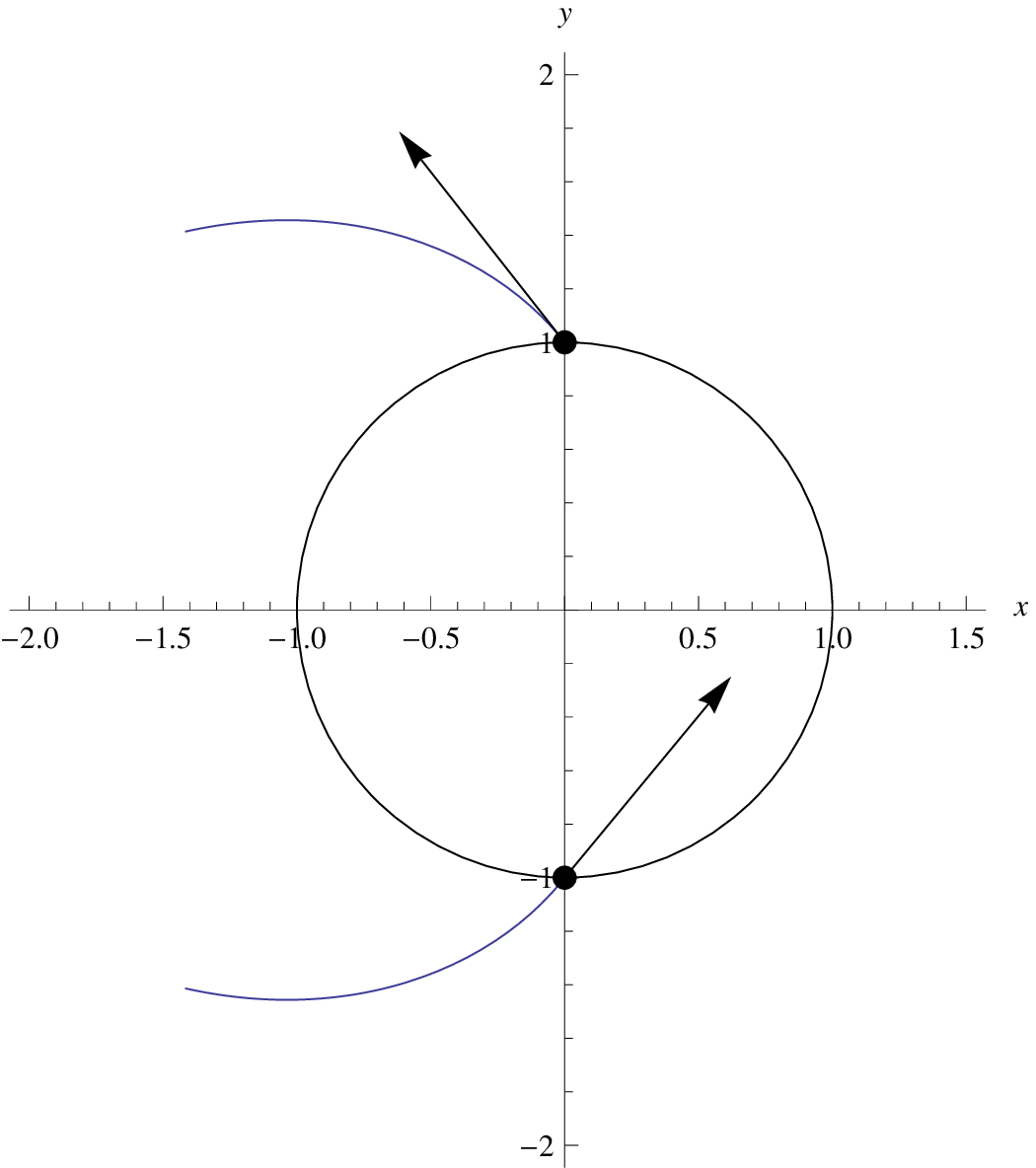}
\hspace*{5mm}
\includegraphics[width=0.35\textwidth]{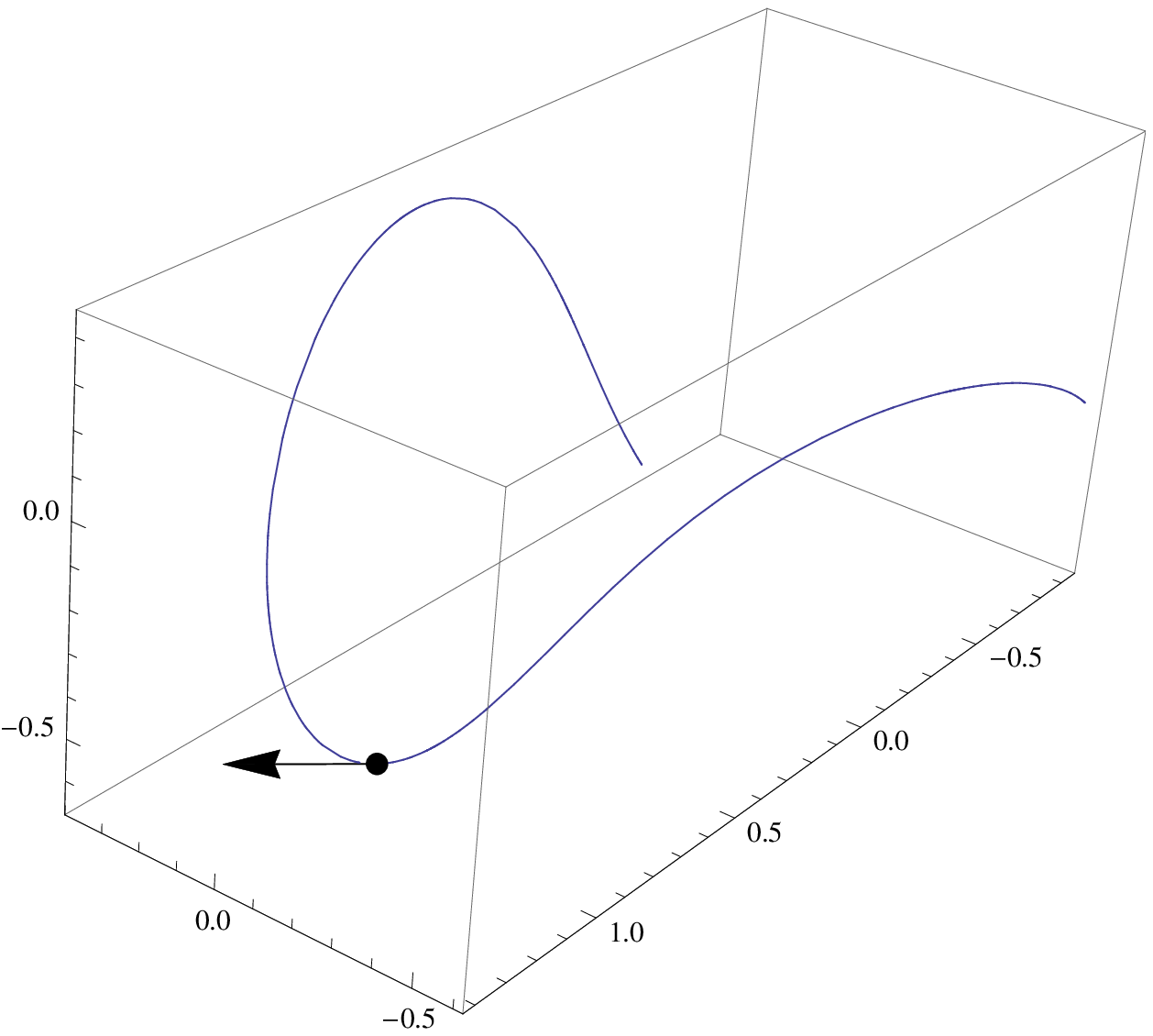}
\end{center}
\caption{Trajectory shown in two representations of $\mathbb{R}P^2$.
Left:   trajectory shown in the 2-plane with surgery.
Right: trajectory shown over Boy's surface (the surface itself is not displayed).
Figures by courtesy of M.~Schwarz\cite{Schwarz2010}.}
\label{fig:Boy's-surface-trajectories}
\vspace{-0cm}
\end{figure}

Two final remarks are in order.
First, the previous view of $ \widetilde{\mathcal{M}}_3$
starting from Euclidean 3-space (Fig.~\ref{fig:defect})
may be partly misleading
if trajectories \emph{through} the defect are considered.
Consider  a particular slice of $\mathbb{R}P^3$ which gives $\mathbb{R}P^2$.
This manifold $\mathbb{R}P^2$, the real projective plane,
can be immersed in $\mathbb{R}^3$ as Boy's surface
(Fig.~\ref{fig:Boy's-surface}).
Now, one particular trajectory
may not look smooth at all in the view based on the
2-plane with surgery (Fig.~\ref{fig:Boy's-surface-trajectories}--left)
but is manifestly smooth if viewed as a curve over
Boy's surface (Fig.~\ref{fig:Boy's-surface-trajectories}--right).

Second, general relativity is all about coordinate
independence of the physics, but a pre-requisite is to have
at least one set of proper coordinates covering
the manifold. Precisely this pre-requisite has been
established with the coordinates \eqref{eq:XnYnZn}.

\section{Nonsingular Defect Solution}
\label{sec:Nonsingular-Defect-Solution}

As explained in the Introduction,
our goal is to find a genuine solution  
of the vacuum Einstein field equations, with a parameter
$b>0$ and topology as suggested by the sketch in Fig.~\ref{fig:defect}
and detailed in Sec.~\ref{sec:Topology}.

A spherically symmetric \textit{Ansatz} for the metric
over $\widetilde{\mathcal{M}}_4$ is given by the line element%
\bsubeqs\label{eq:metric-Ansatz}
\beqa
 ds^2\,\Big|_\text{chart-2}
 &=&
 - \exp\big[2\,\widetilde{\nu}(W)\big] \, dT^2
 + \exp\big[2\,\widetilde{\lambda}(W)\big] \, dY^2
\nonumber\\&&
 +W \left(dZ^2+\sin^2 Z\, dX^2\right)\,,\\[2mm]
W\,\Big|_\text{chart-2}
&\equiv& \zeta^2\,,
\\[2mm]
\zeta\,\Big|_\text{chart-2}
&\equiv& \sqrt{b^2+Y^2}\,,
\eeqa
\esubeqs
with a length parameter $b>0$.
In \eqref{eq:metric-Ansatz}, only the coordinates of the $n=2$ chart
(\ref{eq:X2Y2Z2-ranges}) have been shown,
dropping the suffix `2' on these coordinates. Recall that the
 `radial' coordinate $Y$ takes values in $(-\infty,\,\infty)$,
with positive values of $Y$ on one side of the defect
and negative values on the other side.

With this \textit{Ansatz} and units $G_N=c=1$, 
the following exact solution\cite{KlinkhamerRahmede2013} of
the vacuum Einstein field equations is obtained:%
\bsubeqs\label{eq:vacuum-solution}
\beqa
\label{eq:vacuum-solution-mutilde}
\exp\big[2\,\widetilde{\nu}(W)\big]
&=&
1-2M/\sqrt{W}\;,
\\[2mm]
\label{eq:vacuum-solution-kappatilde}
\exp\big[2\,\widetilde{\lambda}(W)\big]
&=&
\frac{1-b^2/W}{1  -2M/\sqrt{W}}\;,
\eeqa
\esubeqs
in terms of a further parameter $M$ taken to obey
\beq\label{eq:vacuum-solution-defect-parameter}
2M < b\,,
\eeq
so that the metric component \eqref{eq:vacuum-solution-mutilde}
does not vanish over the whole range of $Y$.
The solution for $M=0$ will be seen to correspond to the
flat spacetime with nontrivial topology (\ref{eq:M4})
[see \ref{app:Weakened-Elementary-Flatness-Condition}
for a `blemish' of this spacetime].  
Later, we will comment on
the Schwarzschild-like structure apparent in \eqref{eq:vacuum-solution}.

The resulting Riemann curvature tensor is finite over the whole manifold,
including the defect core at $W=b^2\,$:
\bsubeqs\label{eq:Riemann-defect-sol}
\begin{eqnarray}
R^T_{\ Y T Y}&=&
\frac{2M \left(W-b^2\right)}{W^2 \big(\sqrt{W}-2M \big)}\,,\\[2mm]
R^T_{\ Z T Z}&=& -M/\sqrt{W}\,,\\[2mm]
R^Y_{\ Z Y Z}&=& -M/\sqrt{W}\,,\\[2mm]
R^Z_{\ X Z X}&=& (\sin Z)^2\;2M/\sqrt{W}\,.
\end{eqnarray}
\esubeqs
More importantly, also the Kretschmann curvature scalar,
\begin{eqnarray}\label{eq:Kretschmann-defect-sol}
K &\equiv& R_{\mu\nu\rho\sigma}R^{\mu\nu\rho\sigma}
=48\;\frac{M^2}{W^3}\,,
\end{eqnarray}
remains finite over the whole of $\widetilde{\mathcal{M}}_4$,
because $W \geq b^2 >0$. Some details of the calculation of the
Riemann curvature tensor are presented in \ref{app:Riemann-Curvature-Tensor}.

\begin{figure}[b] 
\begin{center}
\includegraphics[width=0.5\textwidth]{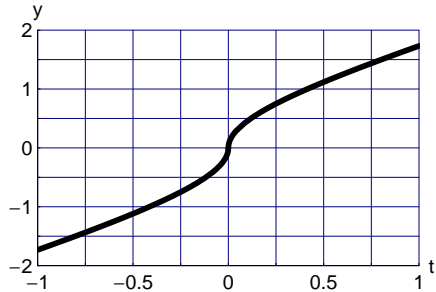}
\end{center}
\caption{Radial geodesic \eqref{eq:radial-geodesics-nonzero-vel}
in terms of dimensionless coordinates $y$ and $t$.}
\label{fig:radial-geodesic}
\vspace{0cm}
\end{figure}

Radial geodesics can be readily calculated.\cite{KlinkhamerRahmede2013}
It is even possible to obtain explicit solutions
for the special case $M=0$ (flat spacetime). Up to arbitrary time shifts,
the radial geodesics are then given in terms of two real constants
$A$ and $B$ (with $B$ taken positive):
\bsubeqs\label{eq:radial-geodesics}
\beqa\label{eq:radial-geodesics-zero-vel}
Y(T) &=& A\;b\,,
\\[2mm]\label{eq:radial-geodesics-nonzero-vel}
Y(T) &=& \left\{\begin{array}{ll}
                \pm\,\sqrt{(B\, T)^2+2\, B\, T}\;\,b&\quad\text{for}\quad T \geq 0\,,
                \\[2mm]
                \mp\,\sqrt{(B\, T)^2-2\, B\, T}\;\,b&\quad\text{for}\quad T < 0\,,
              \end{array}\right.
\eeqa
\esubeqs
where the lower entries in front of the square roots
on the right-hand side of (\ref{eq:radial-geodesics-nonzero-vel})
correspond to a negative asymptotic velocity
and the upper entries to a positive asymptotic velocity
(Fig.~\ref{fig:radial-geodesic}).

The spacetime manifold $\widetilde{\mathcal{M}}_4$ with topology \eqref{eq:M4-M3}
is covered by three coordinate charts,
as explained in Sec.~\ref{sec:Topology}
and \ref{app:Coordinate-Charts}.
Using the results \eqref{eq:metric-Ansatz} and \eqref{eq:vacuum-solution}
for the chart-2 coordinates  (temporarily restoring the suffix `2'),
the respective line elements are
\bsubeqs\label{eq:ds2-defect-charts-123}
\beqa\label{eq:ds2-defect-charts-1}
\hspace*{-0mm}
ds^2\,\Big|_\text{chart-1}
&=&
- \left(1-\frac{2 M}{\sqrt{b^2+(X_{1})^2}}\right)\;dT^2
\nonumber\\&&
+  \left(1-\frac{2 M}{\sqrt{b^2+(X_{1})^2}}\right)^{-1} \;
\frac{(X_{1})^2}{b^2+(X_{1})^2}\;\;(dX_{1})^2
\nonumber\\&&
+ \Big(b^2+(X_{1})^2\Big)\, \Big( (dZ_{1})^2
+ \big(\sin Z_{1}\big)^2\; (dY_{1})^2 \Big)\,,
\eeqa
\beqa
\label{eq:ds2-defect-charts-2}
\hspace*{-0mm}
ds^2\,\Big|_\text{chart-2}
&=&
- \left(1-\frac{2 M}{\sqrt{b^2+(Y_{2})^2}}\right)\;dT^2
\nonumber\\&&
+  \left(1-\frac{2 M}{\sqrt{b^2+(Y_{2})^2}}\right)^{-1} \;
\frac{(Y_{2})^2}{b^2+(Y_{2})^2}\;\;(dY_{2})^2
\nonumber\\&&
+ \Big(b^2+(Y_{2})^2\Big)\, \Big( (dZ_{2})^2
+ \big(\sin Z_{2}\big)^2\; (dX_{2})^2 \Big)\,,
\eeqa
\beqa
\hspace*{-0mm}
\label{eq:ds2-defect-charts-3}
ds^2\,\Big|_\text{chart-3}
&=&
- \left(1-\frac{2 M}{\sqrt{b^2+(Z_{3})^2}}\right)\;dT^2
\nonumber\\&&
+  \left(1-\frac{2 M}{\sqrt{b^2+(Z_{3})^2}}\right)^{-1} \;
\frac{(Z_{3})^2}{b^2+(Z_{3})^2}\;\;(dZ_{3})^2
\nonumber\\&&
+ \Big(b^2+(Z_{3})^2\Big)\, \Big( (dY_{3})^2
+ \big(\sin Y_{3}\big)^2\; (dX_{3})^2 \Big)\,.
\eeqa
\esubeqs
This completes the discussion of the
nonsingular defect solution with parameters $b>0$ and $2M<b$.

\section{Nonsingular Black-Hole Solution}
\label{sec:Nonsingular-BH-Solution}

The defect metric
(\ref{eq:metric-Ansatz})--(\ref{eq:vacuum-solution}) can be
seen to take \emph{precisely} the form
of the standard exterior-region Schwarzschild solution,
in line with Birkhoff's theorem\cite{HawkingEllis1973,MTW1974}.
If $Y^2+b^2$ is identified with $r^2$, the line element
from \eqref{eq:metric-Ansatz} and \eqref{eq:vacuum-solution}
contains the term $(1-b^2/W)\,dY^2 = Y^2/(b^2+Y^2)\,dY^2= dr^2$
and the line element indeed takes the standard Schwarzschild form.

Starting from this observation, we can obtain
a black-hole solution\cite{Klinkhamer2013-MPLA} for the topology
\bsubeqs
\beqa\label{eq:M4-again}
\widetilde{\mathcal{M}}_4
&=&
\mathbb{R} \times \widetilde{\mathcal{M}}_3\,,
\\[2mm]
\widetilde{\mathcal{M}}_3
&\simeq&
\mathbb{R}P^3 - \{\text{point}\}\,,
\eeqa
\esubeqs
with parameters:
\beq
2M > b >0\,.
\eeq
For this new black-hole solution (and another one
described in Sec.~\ref{sec:Nonsingular-BH-Solution-with-Electric-Charge}),
the curvature singularity
will be eliminated by a spacetime defect, i.e., a ``hole'' in spacetime.

The solution will be presented in terms of two sets of
chart-2 coordinates, one set appropriate to the spacetime defect
and another set further out (here, taken to be Kruskal--Szekeres-type
coordinates\cite{Kruskal1960,Szekeres1960}).
But it is also possible to use only one set of chart-2 coordinates; see
\ref{app:Nonsingular-solution-in-Different-coordinates} for details.

At and near the spacetime defect, the metric is given by
\bsubeqs\label{eq:new-BH-solution-TYcoord-ds2-defect-included}
\beqa
ds^2\,\Big|_\text{chart-2}^{(b \leq\,  \zeta < 2M)}
&=&
+ \left(\frac{2 M}{\zeta}-1\right)\;dT^2
-  \left(\frac{2 M}{\zeta}-1\right)^{-1} \;
\frac{Y^2}{\zeta^2}\;\;dY^2
\nonumber\\
&&
+ \zeta^2\, \Big( dZ^2 +\sin^2Z\; dX^2 \Big)\,,
\\[2mm]
\zeta(Y)\,\Big|_\text{chart-2} &=& \sqrt{b^2+Y^2}\,.
\eeqa
\esubeqs
Further out, the metric is given by
\bsubeqs\label{eq:new-BH-solution-TYcoord-ds2-defect-excluded}
\beqa\label{eq:new-BH-solution-TYcoord-ds2-defect-excluded-ds2}
ds^2\,\Big|_\text{chart-2}^{(b <\,  \zeta)}
&=&
32\, M^3\; \frac{\exp\big[-\zeta/(2M)\big]}{\zeta}\;
\Big( -d V^2 +d U^2\Big)
\nonumber\\
&&
+ \zeta^2\, \Big( dZ^2 +\sin^2Z\; dX^2 \Big)\,,
\eeqa
with $\zeta$ written in terms of
Kruskal--Szekeres-type coordinates $U,\,V\in \mathbb{R}$  by
use of the principal branch of the Lambert $W$--function:
\beq\label{eq:new-BH-solution-TYcoord-ds2-defect-excluded-zeta}
\zeta(U,\,V)\,\Big|_\text{chart-2} =
2M \left( 1 + W_{0}\left[\frac{U^2 -V^2}{e}\right]\right)\,.
\eeq
\esubeqs
Recall that the principal branch of the
Lambert $W$--function,  
$W_{0}[z]$, gives the principal solution
for $w$ in \mbox{$z=w \,\exp[w]$} and
that $W_{0}[x]$ is real for $x \in \mathbb{R}$ and $x \geq -1/e$.
The radial coordinate $\zeta$ from
(\ref{eq:new-BH-solution-TYcoord-ds2-defect-excluded-zeta})
reduces to the following expression:
\bsubeqs\label{eq:new-BH-solution-coordinate-zeta-U-V}
\beq\label{eq:new-BH-solution-coordinate-zeta}
\zeta(U,\,V)\,\Big|_\text{chart-2} =\sqrt{b^2+Y^2}\,,
\eeq
by use of the coordinate transformations\cite{HawkingEllis1973,MTW1974}
\begin{eqnarray}\label{eq:new-BH-solution-coordinate-U}
U\,\Big|_\text{chart-2}  &=&
\left\{
\begin{array}{c}
\left(\fracnew{\sqrt{\textstyle{Y^2+b^2}}}{\textstyle{2M}}-1\right)^{1/2}\;
\exp\left[\fracnew{\sqrt{\textstyle{Y^2+b^2}}}{\textstyle{4M}}\right]\,
\cosh\left[\fracnew{\textstyle{T}}{\textstyle{4M}}\right] \,,
 \\[5mm]
\left(1-\fracnew{\sqrt{\textstyle{Y^2+b^2}}}{\textstyle{2M}}\right)^{1/2}\;
\exp\left[\fracnew{\sqrt{\textstyle{Y^2+b^2}}}{\textstyle{4M}}\right]\,
\sinh\left[\fracnew{\textstyle{T}}{\textstyle{4M}}\right]  \,,
\end{array}
\right.
\\[4mm]
\label{eq:new-BH-solution-coordinate-V}
V\,\Big|_\text{chart-2}
&=&
\left\{
\begin{array}{c}
\left(\fracnew{\sqrt{\textstyle{Y^2+b^2}}}{\textstyle{2M}}-1\right)^{1/2}\;
\exp\left[\fracnew{\sqrt{\textstyle{Y^2+b^2}}}{\textstyle{4M}}\right]\,
\sinh\left[\fracnew{\textstyle{T}}{\textstyle{4M}}\right]\,,
 \\[5mm]
\left(1-\fracnew{\sqrt{\textstyle{Y^2+b^2}}}{\textstyle{2M}}\right)^{1/2}\;
\exp\left[\fracnew{\sqrt{\textstyle{Y^2+b^2}}}{\textstyle{4M}}\right]\,
\cosh\left[\fracnew{\textstyle{T}}{\textstyle{4M}}\right]\,,
\end{array}
\right.
\end{eqnarray}
\esubeqs
with top entries for the exterior region \big($\sqrt{Y^2+b^2} > 2M$\big) and
bottom entries for the interior region \big($\sqrt{Y^2+b^2} < 2M$\big).

The Riemann tensor takes the same form as for the
defect solution, see \eqref{eq:Riemann-defect-sol}.
The Ricci tensor and Ricci scalar vanish identically.
Hence, the vacuum Einstein field equations are solved.
Furthermore, the Kretschmann scalar is given by
\begin{eqnarray}\label{eq:Kscalar-new-solution}
K  &\equiv& R_{\mu\nu\rho\sigma}\,R^{\mu\nu\rho\sigma}
=48\;\frac{M^2}{\zeta^6}\,,
\end{eqnarray}
with $\zeta^2 = b^2+Y^2$ over the chart-2 domain, both for the inner
metric (\ref{eq:new-BH-solution-TYcoord-ds2-defect-included})
and the outer metric (\ref{eq:new-BH-solution-TYcoord-ds2-defect-excluded}),
and similarly over the two other domains.
The Kretschmann scalar remains finite because $b>0$.

Recall that the standard Schwarzschild--Kruskal--Szekeres
solution\cite{Schwarzschild1916,Kruskal1960,Szekeres1960}
with topology
\beq
\mathcal{M}_\text{SKS} =\mathbb{R}^2 \times S^2\,,
\eeq
has a physical singularity for $r\to 0$,
as shown by the divergence of the Kretschmann scalar,
\begin{eqnarray}\label{eq:Kscalar-SKS-solution}
K\;\Big|_\text{SKS}
&\equiv&
R_{\mu\nu\rho\sigma}\,R^{\mu\nu\rho\sigma}\;\Big|_\text{SKS}
=48\;\frac{M^2}{r^6}\,.
\end{eqnarray}

The comparison of \eqref{eq:Kscalar-new-solution} for $b\ne 0$
and \eqref{eq:Kscalar-SKS-solution} makes clear that the solution
\eqref{eq:new-BH-solution-TYcoord-ds2-defect-included}%
--\eqref{eq:new-BH-solution-TYcoord-ds2-defect-excluded}
over $\mathbb{R} \times \widetilde{\mathcal{M}}_3$
may be considered to be a \emph{regularized} version
of the standard Schwarzschild solution
over  $\mathbb{R}^{+} \times \mathbb{R} \times S^2$,
with the curvature singularity
eliminated by a spacetime defect, i.e., a ``hole'' in spacetime.

The coordinate $T$ and the coordinate $Y$
are, respectively, spacelike and timelike
for the inner metric (\ref{eq:new-BH-solution-TYcoord-ds2-defect-included}).
This behavior is analogous
to what happens for the standard Schwarzschild solution\cite{MTW1974}.
Note, however, that the timelike coordinate $Y$ of the
inner metric (\ref{eq:new-BH-solution-TYcoord-ds2-defect-included})
ranges from $-\infty$ to $+\infty$, unlike the usual radial coordinate $r$.
Moreover, this timelike coordinate $Y$ is part of a topologically
nontrivial manifold $\widetilde{\mathcal{M}}_3$.
This gives rise to the presence of closed time-like curves (CTCs).
These CTCs imply all possible horrors, but, classically,
these horrors remain \emph{confined} within the Schwarzschild horizon.
Whether or not CTCs in the interior region
are physically acceptable depends on the behavior of the matter
fields.\footnote{\label{ftn:reg-BH-causality}%
Remark that the regular black-hole solution
from nonlinear electrodynamics\cite{AyonBeatoGarcia1998,Garcia-etal2013}
may also have problems with causality:
absence of a global Cauchy surface
(i.e., incomplete predictability of the future)
and possible lack of microcausality for the electromagnetic fields
(cf. Refs.~\citen{KosteleckyLehnert2000,AdamKlinkhamer2001}).}

\section{Nonsingular Black-Hole Solution with Electric Charge}
\label{sec:Nonsingular-BH-Solution-with-Electric-Charge}

The problematic closed timelike curves   
of the modified Schwarzschild solution
(\ref{eq:new-BH-solution-TYcoord-ds2-defect-included}) trace
back to the fact that the original singularity
was \emph{spacelike}. But it is well-known
that the singularity of the standard Reissner--Nordstr\"{o}m (RN)
solution\cite{Reissner1916,Nordstrom1916}
is  \emph{timelike}\cite{GravesBrill1960,Carter1966}
(see also Refs.~\citen{HawkingEllis1973,MTW1974}).
This suggest, first, to
add a small electric charge and, then, to modify
the resulting RN
solution in order to arrive at a nonsingular solution\cite{Klinkhamer2013-APPB}.

Consider, then, spherically symmetric solutions
of the Einstein field equations,
\bsubeqs\label{eq:Einstein-equations-T-ext}  
\beqa\label{eq:Einstein-equations}
R_{\mu}^{\;\;\nu}(x) - \frac{1}{2}\, R(x)\, \delta_{\mu}^{\;\;\nu}
&=&
8\pi\,T_{\mu}^{\;\;\nu}(x) \,,
\eeqa
where the energy-momentum tensor $T_{\mu}^{\;\;\nu}$
is set equal to a \emph{prescribed} energy-momentum
tensor $\Theta_{\mu}^{\;\;\nu}$ (using spherical coordinates),
\beqa\label{eq:T-ext}
\hspace*{-4mm}
T_{\mu}^{\;\;\nu}(t,\,r,\,\theta,\,\phi)
&=&
\Theta_{\mu}^{\;\;\nu}(t,\,r,\,\theta,\,\phi)
\equiv
\frac{Q^2}{8\pi\,r^4}\;
D_{\mu}^{\;\;\nu}(t,\,r,\,\theta,\,\phi)\,,
\eeqa
in terms of the traceless diagonal matrix
\beqa
D_{\mu}^{\;\;\nu}(t,\,r,\,\theta,\,\phi)
&\equiv&
\left\{\begin{array}{cl}
-1 & \;\;\text{for}\;\;\mu=\nu\in \{t,\,r\}\,,\\
+1 & \;\;\text{for}\;\;\mu=\nu\in \{\theta,\,\phi\}\,,\\
0  & \;\;\text{otherwise}\,.
\end{array}\right.
\eeqa
\esubeqs
This particular $\Theta_{\mu}^{\;\;\nu}$
corresponds to the energy-momentum tensor of a Coulomb-type electric field.
(It is also possible to deal with the coupled Einstein and Maxwell equations,
but we simplify the discussion by use of a fixed Coulomb-type
energy-momentum tensor.)

The standard Reissner--Nordstr\"{o}m
solution in the exterior region has a metric given by  the following line element:
\beqa\label{eq:RN-solution}
\hspace*{-0mm}
ds^2\,\Big|^{(r>r_{+})}_\text{RN}
&=&
-\left(1- \frac{2 M}{r} + \frac{Q^2}{r^2}\right)\; dt^2
+\left(1- \frac{2 M}{r} + \frac{Q^2}{r^2}\right)^{-1}\; dr^2
\nonumber\\
&&
+ r^2\, \Big( d\theta^2 +\sin^2\theta\; d\phi^2 \Big)\,,
\eeqa
with coordinates $t \in \mathbb{R}$,
$r>r_{+}\equiv M + \sqrt{M^2-Q^2}$, $\theta\in [0,\,\pi]$,
$\phi\in [0,\,2\pi)$.
Here, $M$ can be interpreted as the mass of the central object
and $Q$ as its electric charge.

The corresponding nonsingular solution in terms of an
effective radial coordinate $\zeta$
will be seen to have a further length parameter $b$.
The three parameters of the solution are taken to be related as follows:
\bsubeqs\label{eq:conditions-QM-b}
\beqa\label{eq:conditions-QM}
0 \; < \; &|Q|&  \;<\;  M\,,
\\[2mm]
\label{eq:conditions-b}
0 \; <\; &b&     \;<\;  \zeta_{-}\,,
\eeqa
\esubeqs
with definitions
\beqa\label{eq:beta-pm}
\zeta_{\pm} &\equiv& M \pm \sqrt{M^2-Q^2}\,.
\eeqa
Note that, for the classical theory,
the electric charge $|Q|$ can be arbitrarily small,
as long as the charge $Q$ remains nonzero
and the length $b$ is sufficiently small,
\mbox{$b < \zeta_{-} \sim Q^2/(2M)$.}
[A different choice of parameters is considered in
Footnote~\ref{ftn:charged-BH-PG} of
\ref{app:Nonsingular-solution-in-Different-coordinates}.]

For the construction of the nonsingular solution
with parameters (\ref{eq:conditions-QM-b}), we refer
to Carter's original article\cite{Carter1966}
for the standard Reissner--Nordstr\"{o}m solution
and follow the same modification procedure  used
in Sec.~\ref{sec:Nonsingular-BH-Solution}
for the Schwarzschild solution.
As we are primarily interested in the removal of the
curvature singularity, we focus on the spacetime
region III ($0 \leq \zeta < \zeta_{-}$).
No essential change occurs for the spacetime regions I and II
($\zeta \geq \zeta_{-}>0$),
because they do not reach the singularity at $\zeta=0$.

The region--III metric with the defect at $\zeta=b$ included
is then found to have the following line element:
\bsubeqs\label{eq:new-solution-region-III}
\beqa\label{eq:new-solution-region-III-ds2}
\hspace*{-0mm}
ds^2\,\Big|^{(b \leq\zeta < \zeta_{-})}_\text{chart-2}
&=&
-\left(1- \frac{2 M}{\zeta} + \frac{Q^2}{\zeta^2}\right)\; dT^2
+\left(1- \frac{2 M}{\zeta} + \frac{Q^2}{\zeta^2}\right)^{-1}\;
\frac{Y^2}{\zeta^2}\;dY^2
\nonumber\\
&&
+ \zeta^2\, \Big( dZ^2 +\sin^2 Z\; dX^2 \Big)\,,
\\[2mm]
\label{eq:new-solution-region-III-zeta}
\zeta\,\Big|_\text{chart-2}
&=& \sqrt{b^2+Y^2}\,.
\eeqa
\esubeqs
From the standard analysis\cite{HawkingEllis1973,MTW1974},
it follows that the singularities at
$\zeta=\zeta_{\pm}$ in (\ref{eq:new-solution-region-III-ds2})
can be removed by coordinate transformations.

Note that (\ref{eq:new-solution-region-III-ds2})
takes precisely the form of the original Reissner-Nordstr\"{o}m metric (\ref{eq:RN-solution})
if $(Y^2/\zeta^2)\,dY^2$ is replaced by $d\zeta^2$
according to (\ref{eq:new-solution-region-III-zeta}).
But, as emphasized before, the crucial point here is the appearance of
the coordinate $Y \in (-\infty,\,\infty)$
of the nonsimply-connected manifold $\widetilde{\mathcal{M}}_3$.
In addition, there are now radial geodesics passing through $Y=0$, as
explained in the penultimate paragraph of
Sec.~\ref{sec:Nonsingular-Defect-Solution}.

The metric (\ref{eq:new-solution-region-III}) solves
the Einstein field equations (\ref{eq:Einstein-equations})
for a  prescribed  energy-momentum tensor $\Theta_{\mu}^{\;\;\nu}(T,\,X,\, Y,\, Z)$
of the diagonal form (\ref{eq:T-ext})
with $1/r^4$ replaced by $1/\zeta^4=1/(b^2+Y^2)^2$.

The spacetime from the metric (\ref{eq:new-solution-region-III}),
extended to all charts, corresponds to a
noncompact, orientable, nonsimply-connected manifold without boundary.
This spacetime has the topology%
\bsubeqs\label{eq:new-solution-region-III-Mtilde4-Mtilde3}
\beqa\label{eq:new-solution-region-III-Mtilde4}
\widetilde{\mathcal{M}}\,\Big|_\text{mod-RN}
 &=& \mathbb{R} \times \widetilde{\mathcal{M}}_3 \,,
 \\[2mm]
\label{eq:new-solution-region-III-Mtilde3}
\widetilde{\mathcal{M}}_3
&\simeq&
\mathbb{R}P^3 - \{\text{point}\}\,,
\eeqa
\esubeqs
where `mod-RN' stands for the modified Reissner-Nordstr\"{o}m solution and
$\mathbb{R}P^3$ is the 3-dimensional real projective space
(topologically equivalent to a 3-sphere with antipodal points identified).

The Kretschmann curvature scalar over the
manifold (\ref{eq:new-solution-region-III-Mtilde4-Mtilde3}) is given by
\begin{eqnarray}\label{eq:metric-Ansatz-Kscalar}
K\,\Big|_\text{mod-RN}
&\equiv&
R_{\mu\nu\rho\sigma}\,R^{\mu\nu\rho\sigma}\,\Big|_\text{mod-RN}
=\frac{8 \,\big(6\, M^2\, \zeta^2- 12\, M\, Q^2\, \zeta + 7\, Q^4\big)}
      {\zeta^8}\,,
\end{eqnarray}
which remains finite because $\zeta >0$ for $b>0$.
For fixed values of $M$ and $Q$
obeying condition (\ref{eq:conditions-QM}),
$K(\zeta)$ drops monotonically with $\zeta$.
This fact allows for an operational definition of $b$
from the maximum value of $K$. (The operational definitions
of $M$ and $Q$  rely, for example, on the asymptotic $\zeta\to\infty$
behavior of the metric and electromagnetic fields.)

The main result of this section is that the
factor $\mathbb{R}$ in (\ref{eq:new-solution-region-III-Mtilde4})
corresponds to the
\emph{timelike} direction of
the metric (\ref{eq:new-solution-region-III}),
making for a \emph{spacelike} hypersurface
$\widetilde{\mathcal{M}}_3$ in the spacetime region III. In turn,
this observation implies the \emph{absence} of closed timelike curves.
Recall that the spacetime regions II ($\zeta_{-}< \zeta < \zeta_{+}$) 
and I ($\zeta >\zeta_{+}$) do not reach the $\zeta=b$ surface 
where antipodal points are identified (cf. Fig.~\ref{fig:defect}).

Mathematically speaking, the nonsingular black-hole
solution (\ref{eq:new-solution-region-III})
with parameters (\ref{eq:conditions-QM-b})
can be viewed as  a ``regularization''  of the singular
Reissner--Nordstr\"{o}m solution.\footnote{\label{ftn:elementary-particles}%
It is well-known that the standard Reissner--Nordstr\"{o}m
metric \eqref{eq:RN-solution} with $M=0$ has a naked singularity.
Not so for the metric \eqref{eq:new-solution-region-III} with $M=0$.
This may provide new impetus to discussions of elementary particles
(rotation/spin neglected, for the moment) being interpreted as tiny
spacetime structures (cf. Sec.~5.2 of Ref.~\citen{Visser1995}).
Note that macroscopic naked singularities
(and their regulated versions) have distinctive gravitational-lensing
characteristics~\cite{VirbhadraEllis2002}.}

\section{Discussion}
\label{sec:Discussion}

Apart from the mathematical interest of having
a new type of exact solution of the Einstein field equations,
these nonsingular solutions (with or without electric charge)
may also appear in a physical context.
Let us focus on the charged nonsingular solution
(\ref{eq:new-solution-region-III})
and perform the following \textit{Gedankenexperiment}.  

Start from a nearly flat spacetime with
a trivial topology $\mathbb{R}^4$ and a
metric approximately equal to the Minkowski metric.
Next, arrange for  a large amount of matter with total mass
$\overline{M}$ and with
vanishing net charge $\overline{Q}=0$
to collapse in a spherically symmetric way.
Within the realm of classical Einstein gravity, 
we expect to end up with the singular Schwarzschild solution. 
But, very close to the final curvature singularity,
something else may happen due to quantum effects.

Consider a precursor mass
$\overline{\Delta M} \sim \hbar/(b\,c) \ll \overline{M}$
and use typical curvature values from the expressions
\eqref{eq:Kscalar-new-solution} and \eqref{eq:Kscalar-SKS-solution}
for the Kretschmann scalar. Then,
the local spacetime integral of the action density  related to the
standard Schwarzschild solution
differs from that related to (\ref{eq:new-solution-region-III})
for $Q^2 \ll M^2$
by an amount $\lesssim \hbar$. As argued by Wheeler in particular,
the local topology of the manifold may now change by a quantum jump
if $b$ is sufficiently close to the length scale
$L_\text{Planck}$ $\equiv$ $(\hbar\, G_N/c^3)^{1/2}$.
In addition, the strong gravitational fields may lead to electron-positron
pair creation, possibly with one charge expelled towards spatial infinity.

These two quantum processes combined may
result in a transition from the simply-connected
manifold $\mathbb{R}^4$ without localized charge $Q$ to the
nonsimply-connected manifold $\widetilde{\mathcal{M}}_4$ with localized charge
$Q$ $=$ $\pm\, |Q_\text{electron}|$ $\equiv$ $\pm\, e$.
Hence, \emph{if}
the transition amplitude between the different topologies
is nonzero for appropriate matter content,
quantum mechanics can operate a change between
the classical Schwarzschild solution
and the classical solution (\ref{eq:new-solution-region-III}) with
$\overline{Q}=\pm\, e$ and an additional charge $\mp\, e$ at infinity.
In this way, the curvature singularity would be eliminated,
but not at the price of introducing closed timelike curves.

It is also possible to present an
alternative scenario \emph{without} topology change.
Now, the spherical collapse of matter is  assumed to
occur in Minkowski spacetime with a relatively sparse
sprinkling of massless static defects
(each one given by the solution of Sec. 2 with $M=0$ and $b>0$).
Then, the precursor mass $\overline{\Delta M}$ selects one of the available
defect cores and
increases its mass
($M=0 \to \overline{\Delta M} \to \overline{M}$),
possibly giving it also a charge
by electron-positron pair creation with one charge expelled to infinity.
Again, the curvature singularity would be eliminated,
while avoiding closed timelike curves.

Many questions remain, the most important of which are the following:
\begin{romanlist}[(iii)]
\item
Are these regularized Schwarzschild solutions
\emph{really} acceptable, both mathematically and physically?    
\vspace*{1mm}\item
Are there perhaps other surprises from this regularization,
in a way reminiscent of the anomalies of quantum field theory? 
\vspace*{1mm}\item
Where does the matter go, is it distributed over a thin shell
with $\zeta \in [b,\, b+\Delta b)$ for positive $\Delta b\,$?%
\vspace*{1mm}\item
Does realistic collapse occur with or without topology change?
\end{romanlist}
\noindent These are obviously difficult questions.
A partial answer to the first question appears
in \ref{app:Weakened-Elementary-Flatness-Condition},
where a particular characteristic of the regularized spacetime
is discussed. Further progress on all questions
can perhaps be made by direct investigations of the matter sector
[cf. Ref.~\citen{Klinkhamer2014} for a nonsingular
defect solution with an $SO(3)$ Skyrmion field].

\section*{Acknowledgments}

It is a pleasure to thank
C.~L\"{a}mmerzahl for discussions on elementary flatness
and A. Macias for pointing out Ref.~\citen{AyonBeatoGarcia1998}.

\appendix
\section{Coordinate Charts}
\label{app:Coordinate-Charts}

The three coordinate charts of the
3-manifold $\widetilde{\mathcal{M}}_3$ with topology \eqref{eq:M3-topology}
were briefly discussed in Sec.~\ref{sec:Topology}.
Further details will be given  in this appendix.

A relatively simple covering of $\widetilde{\mathcal{M}}_3$
uses three charts of coordinates, labeled by $n=1,2,3$.
Each  chart covers and surrounds part of one of the three Cartesian
coordinate axes but does not intersect the other two axes.
For example, the $n=1$ coordinate chart covers and surrounds
the $|x^1|\geq b$ parts of
the $x^1$ coordinate axis but does not intersect the $x^2$ and $x^3$  axes.
The domains of the chart-1 coordinates consist of two `wedges,'
on both sides of the defect and pierced by the $x^1$ axis;
see Fig.~\ref{fig:defect-charts}--left.

These coordinates 
are denoted $(X_n,\,Y_n,\,Z_n)$, for $n=1,\,2,\,3\,$, and their ranges 
were already given by
\eqref{eq:XnYnZn-ranges} of the main text.
In order to describe the interrelation of the coordinates
$(X_n,\,  Y_n,\, Z_n)$ in the overlap regions,
we express them in terms of the
coordinates of the 3-dimensional Euclidean space $E_3$.
For the latter, we use both standard
and nonstandard spherical coordinates. The standard spherical
coordinates $(r,\, \theta,\, \phi)$ are defined by
\beqa\label{eq:standard-spherical-coord}
(x^1,\,  x^2,\, x^3)
&=&
(r \sin\theta  \cos\phi,\,r \sin\theta  \sin\phi,\, r \cos\theta )\,,
\eeqa
with $r \geq 0$, $\theta \in [0,\,\pi]$, and $\phi \in [0,\,2\pi)$.
The nonstandard spherical coordinates $(r,\, \vartheta,\, \varphi)$
are defined by
\beqa\label{eq:nonstandard-spherical-coord}
(x^1,\,  x^2,\, x^3)
&=&
(r \sin\vartheta \sin\varphi,\,r \cos\vartheta,\, r \sin\vartheta\cos\varphi)\,,
\eeqa
with $r \geq 0$, $\vartheta \in [0,\,\pi]$, and $\varphi \in [0,\,2\pi)$.

Now, the chart-1 and chart-2 coordinates over
the appropriate regions (wedges) of $\widetilde{M}_{3}$ are given by
\bsubeqs\label{eq:X1Y1Z1-def}
\beqa
X_{1} &=&
\left\{\begin{array}{ll}
r-b  \hspace*{6.5mm} &\quad\text{for}\quad  \cos\phi > 0\,,\\
b-r       &\quad\text{for}\quad  \cos\phi <    0\,,
\end{array}\right.\\[2mm]
Y_{1} &=&
\left\{\begin{array}{ll}
\phi-\pi/2  &\quad\text{for}\quad  \pi/2 < \phi< 3\pi/2\,,\\
\phi-3\pi/2 &\quad\text{for}\quad  3\pi/2 < \phi< 2\pi\,,\\
\phi+\pi/2  &\quad\text{for}\quad  0 \leq \phi < \pi/2\,,
\end{array}\right.\\[2mm]
Z_{1} &=& \left\{\begin{array}{ll}
\theta       &\quad\text{for}\quad  \cos\phi > 0\,,\\
\pi-\theta\hspace*{5.0mm}   &\quad\text{for}\quad  \cos\phi <    0\,,
              \end{array}\right.
\eeqa
\esubeqs
and
\bsubeqs\label{eq:X2Y2Z2-def}
\beqa
X_2 &=& \left\{\begin{array}{ll}
\phi       &\quad\text{for}\quad  0 < \phi< \pi\,,\\
\phi-\pi   &\quad\text{for}\quad  \pi < \phi < 2\pi\,,
              \end{array}\right.\\[2mm]
Y_2 &=& \left\{\begin{array}{ll}
r-b   \hspace*{1.0mm}    &\quad\text{for}\quad  0 < \phi< \pi\,,\\
b-r       &\quad\text{for}\quad  \pi < \phi < 2\pi\,,
              \end{array}\right.\\[2mm]
Z_2 &=& \left\{\begin{array}{ll}
\theta       &\quad\text{for}\quad  0 < \phi< \pi\,,\\
\pi-\theta   &\quad\text{for}\quad  \pi < \phi < 2\pi\,,
              \end{array}\right.
\eeqa
\esubeqs
in terms of the standard spherical
coordinates \eqref{eq:standard-spherical-coord}.

The $n=3$ chart requires coordinates of $E_3$
that are regular on the Cartesian $x^3$ axis.
The chart-3 coordinates over the relevant regions (wedges)
of $\widetilde{M}_{3}$ are then given by%
\bsubeqs\label{eq:X3Y3Z3-def}
\beqa
X_{3} &=&
\left\{\begin{array}{ll}
\varphi-\pi/2  &\quad\text{for}\quad  \pi/2 < \varphi< 3\pi/2\,,\\
\varphi-3\pi/2 &\quad\text{for}\quad  3\pi/2 < \varphi< 2\pi\,,\\
\varphi+\pi/2  &\quad\text{for}\quad  0 \leq \varphi < \pi/2\,,
\end{array}\right.\\[1mm]
Y_{3} &=&
\left\{\begin{array}{ll}
\vartheta       &\quad\text{for}\quad  \cos\varphi > 0\,,\\
\pi-\vartheta \hspace*{5.0mm}   &\quad\text{for}\quad  \cos\varphi <    0\,,
\end{array}\right.  \\[1mm]
Z_{3} &=&
\left\{\begin{array}{ll}
r-b  \hspace*{6.5mm}     &\quad\text{for}\quad  \cos\varphi > 0\,,\\
b-r       &\quad\text{for}\quad  \cos\varphi <    0\,,
\end{array}\right.
\eeqa
\esubeqs
in terms of the nonstandard spherical coordinates
\eqref{eq:nonstandard-spherical-coord}.

It can be verified that these three sets of coordinates
$(X_n,\,  Y_n,\, Z_n)$ are invertible
and infinitely-differentiable functions of each other
in the overlap regions.
Moreover, the manifold satisfies the Hausdorff property,
i.e., two distinct points can each be surrounded by open sets,
so that these two open sets do not overlap
(see Ref.~\citen{Schwarz2010} for details).

\section{Riemann Curvature Tensor}
\label{app:Riemann-Curvature-Tensor}

In this appendix, 
we calculate the Riemann curvature tensor
for the nonsingular defect metric
\eqref{eq:metric-Ansatz}--\eqref{eq:vacuum-solution}
by using two deformations of the metric.
The deformed metrics never vanish in the domains over which they
are defined and are directly invertible.
The parameters of the defect solution are $b>0$ and $2M<b$,
but the same analysis applies to the case
$2M>b>0$, which corresponds to black-hole-type solutions.

Denote the deformation parameter by $\epsilon$ and assume that
\beq
\epsilon >0\,.
\eeq
The particular deformation considered will be seen to
break the spherical symmetry (e.g., evenness in $Y$).
The limit $\epsilon \to 0$ will be taken
at the end of the calculation.

The two deformed metrics are, in fact, given by
\bsubeqs\label{eq:deformed-metrics}
\beqa
g_{\mu\nu}^{\pm}[T,\, X,\,  Y,\, Z]
&=&
\Big[
\text{diag}
\Big(-\big(1-2M/\zeta^{\pm}\big),\,
(\zeta^{\pm})^2\; \sin^2 Z,\,
\nonumber\\&&
\big(1-2M/\zeta^{\pm}\big)^{-1}\;(d\zeta^{\pm}/d Y)^2,\,
(\zeta^{\pm})^2\Big)\Big]_{\mu\nu}\,,
\\[2mm]
\zeta^{\pm}
&=&
\sqrt{b^2+Y^2 \pm\epsilon\,Y\,b}\,,
\eeqa
\esubeqs
for chart-2 coordinates without suffix `2'.
These two metrics are defined over two overlapping $Y$ domains:
\bsubeqs\label{eq:deformed-metrics-plus-minus}
\beqa
g_{\mu\nu}^{+}
&\;\;\text{for}\;\;& Y \in (-\delta,\,+\infty)\,,
\\[2mm]
g_{\mu\nu}^{-}
&\;\;\text{for}\;\;& Y \in (-\infty,\,+\delta)\,,
\eeqa
with an arbitrary value of $\delta$ in the open range $(0,\,\epsilon\, b/2)$,
for example
\beqa
\delta &=&\epsilon\, b/4\,.
\eeqa
\esubeqs

The Riemann tensors from these metrics are readily calculated
and the corresponding Kretschmann scalars are found to be given by
\beq
K^{\pm}[Y]
=
48 \;\frac{M^2}{\big(b^2+Y^2 \pm\epsilon\,Y\,b\big)^{3}}\,.
\eeq
With these results, we can define
\beq
K[Y]
=
\left\{\begin{array}{lcl}
\lim_{\epsilon\to 0}\,K^{+}[Y]&\;\;\text{for}\;\;&Y \geq 0 \,,\\[1mm]
\lim_{\epsilon\to 0}\,K^{-}[Y]&\;\;\text{for}\;\;&Y < 0 \,,
              \end{array}\right.
\eeq
which agrees with the previous result \eqref{eq:Kretschmann-defect-sol}.
The point of the above  exercise is that $K$ has now been calculated
with nonvanishing (invertible) metrics altogether.

\section{Nonsingular Solution in Different Coordinates}
\label{app:Nonsingular-solution-in-Different-coordinates}

In this appendix, we present the nonsingular black-hole
solution from Sec.~\ref{sec:Nonsingular-BH-Solution} in terms
of a different set of coordinates.
The advantage of these new coordinates
is that we only need one set of chart-2 coordinates and
not two sets as used in
(\ref{eq:new-BH-solution-TYcoord-ds2-defect-included})
and  (\ref{eq:new-BH-solution-TYcoord-ds2-defect-excluded}).

Instead of starting from
an \textit{Ansatz} based on Kruskal--Szekeres coordinates
as was done in our original article\cite{Klinkhamer2013-MPLA},
we start from an \textit{Ansatz} based on Painlev\'{e}--Gullstrand (PG)
coordinates\cite{Painleve1921,Gullstrand1922}
(see, e.g., Ref.~\citen{MartelPoisson2000} for a brief review).
Turning immediately to the chart-1 coordinates
from \ref{app:Coordinate-Charts} and
replacing the radial coordinate $r$ of the PG coordinates by
$\sqrt{b^2+(X_{1})^2}$, we arrive at the following line element
over part of the manifold \eqref{eq:M4-M3}:
\bsubeqs\label{eq:new-BH-solution-PVcoord-chart-123}
\beqa\label{eq:new-BH-solution-PVcoord-chart-1}
ds^2\,\Big|_\text{chart-1}
&=&
- d\widehat{T}^2
+\left(\frac{X_{1}}{\sqrt{b^2+(X_{1})^2}}\;\;d X_{1}
+\sqrt{\frac{2 M}{\sqrt{b^2+(X_{1})^2}}}\;\;d\widehat{T}\right)^2
\nonumber\\&&
+(b^2+(X_{1})^2)\,\Big( (dZ_{1})^2+\big(\sin Z_{1}\big)^2\;(dY_{1})^2\Big)\,,
\eeqa
for mass parameter $M>0$ and length parameter $b>0$.
Remark that the surfaces of constant $\widehat{T}$ are
intrinsically flat\cite{MartelPoisson2000}.
An advantage of the metric \eqref{eq:new-BH-solution-PVcoord-chart-1}
is that it applies not only to
the black-hole case $2M\geq b$ (including the special case $2M=b$)
but also to the defect case $2M < b$.\footnote{\label{ftn:charged-BH-PG}%
Similar PG-type coordinates $(\widehat{T},\, X_{1},\,  Y_{1},\, Z_{1})$
can be used for the nonsingular charged black hole with the
parameter choice $b \geq Q^2/(2M)>0$, which differs from the
choice \eqref{eq:conditions-b} considered in
Sec.~\ref{sec:Nonsingular-BH-Solution-with-Electric-Charge}.
The metric then takes the same form as the one from
\eqref{eq:new-BH-solution-PVcoord-chart-1} but
with $2 M/\sqrt{b^2+(X_{1})^2}$
replaced by $2 M/\sqrt{b^2+(X_{1})^2}-Q^2/(b^2+(X_{1})^2)$.
For fixed positive values of $b$ and $M$, it is now possible
to take the classical charge $|Q|$ arbitrarily small.}

We obtain the metrics for the $n=2$ and $n=3$ charts
by taking the coordinates $(X_2,\,  Y_2,\, Z_2)$ and $(X_3,\,  Y_3,\, Z_3)$
instead of $(X_{1},\,  Y_{1},\, Z_{1})$.
With the same replacements as used in
\eqref{eq:ds2-defect-charts-123}, the corresponding metrics
are given by the following line elements:
\beqa\label{eq:new-BH-solution-PVcoord-chart-2}
ds^2\,\Big|_\text{chart-2}
&=&
- d\widehat{T}^2
+\left(\frac{Y_{2}}{\sqrt{b^2+(Y_{2})^2}}\;\;d Y_{2}
+\sqrt{\frac{2 M}{\sqrt{b^2+(Y_{2})^2}}}\;\;d\widehat{T}\right)^2
\nonumber\\&&
+(b^2+(Y_{2})^2)\,\Big((dZ_{2})^2+(\sin Z_{2})^2\;(dX_{2})^2\Big)\,,
\\[2mm]
\label{eq:new-BH-solution-PVcoord-chart-3}
ds^2\,\Big|_\text{chart-3}
&=&
- d\widehat{T}^2
+\left(\frac{Z_{3}}{\sqrt{b^2+(Z_{3})^2}}\;\;d Z_{3}
+\sqrt{\frac{2 M}{\sqrt{b^2+(Z_{3})^2}}}\;\;d\widehat{T}\right)^2
\nonumber\\&&
+(b^2+(Z_{3})^2)\,\Big((dY_{3})^2+\big(\sin Y_{3}\big)^2\;(dX_{3})^2\Big)\,.
\eeqa
\esubeqs

The PG coordinates play a special role in discussions of emergent gravity
in superfluid systems\cite{Unruh1981,Volovik2009}.
Referring to \eqref{eq:new-BH-solution-PVcoord-chart-123},
it appears, in principle, possible to have
Unruh-type artificial black holes without curvature singularity.
If such artificial black holes can indeed be realized
with superfluids in the laboratory,
the existence of closed time-like curves from the effective metric
(last paragraph of Sec.~\ref{sec:Nonsingular-BH-Solution})
may signal the appearance of new topology-driven instabilities of the
quasi-particle system.

\section{Weakened Elementary-Flatness Condition}
\label{app:Weakened-Elementary-Flatness-Condition}

In this appendix, we take a closer look at the
metric \eqref{eq:new-solution-region-III} near the defect core.
In principle, it is possible to set $Q=0$ and even $M=Q=0$,
as long as $b$ remains nonzero (here, taken to be positive).

Specifically, consider a small neighbourhood around
the spacetime point $P$ with the following chart-2 coordinates:
\beq\label{eq:point-P}
(T,\, X_2,\,  Y_2,\, Z_2)\Big|_{P} =(0,\, 0,\,  0,\, \pi/2)\,.
\eeq
Next, define dimensional coordinates
$(t,\, \widetilde{x},\,y,\, \widetilde{z})$ which vanish at $P$ :
\beq\label{eq:new-coord}
(T,\, X_2,\,  Y_2,\, Z_2) = (0,\,  0,\,  0,\, \pi/2)+
(t,\, \widetilde{x}/b,\,y,\, \widetilde{z}/b)\,.
\eeq
The metric \eqref{eq:new-solution-region-III}
near $P$  then becomes
\beqa\label{eq:metric-near-P}
ds^2\,\Big|_{\text{near-}P}
&=&
-\left(1- \frac{2 M}{\sqrt{b^2+y^2}} + \frac{Q^2}{b^2+y^2}\right)\; dt^2
\nonumber\\
&&
+\left(1- \frac{2 M}{\sqrt{b^2+y^2}} + \frac{Q^2}{b^2+y^2}\right)^{-1}\;
\frac{y^2}{b^2+y^2}\;dy^2
\nonumber\\
&&
+ (1+y^2/b^2)\, \Big( d\widetilde{z}^2 +\big[1+O(\widetilde{z}^2/b^2)\big]\;
d\widetilde{x}^2 \Big)
\nonumber\\[2mm]
&\sim&
-\left(1- 2 M/b + Q^2/b^2\right)\; dt^2
\nonumber\\
&&
+\left(1- 2 M/b + Q^2/b^2\right)^{-1}\;
\frac{y^2}{b^2}\;dy^2
+ d\widetilde{z}^2 +d\widetilde{x}^2 \,.
\eeqa
With a further change of coordinates,
\bsubeqs\label{eq:new-coord-final}
\beqa\label{eq:new-coord-final-t-tilde}
\widetilde{t}
&=&
A\;t \,,
\\[2mm]
\label{eq:new-coord-final-y-tilde}
\widetilde{y}
&=&
\left\{
\begin{array}{ccl}
+y^2/(2b\,A) &\;\text{for}\;& y\geq 0\,,
 \\[2mm]
-y^2/(2b\,A) &\;\text{for}\;& y< 0\,,
\end{array}
\right.
\\[2mm]
A &\equiv& \sqrt{1- 2 M/b + Q^2/b^2}\; > \;0\,,
\eeqa
\esubeqs
the metric near $P$  reads
\beq\label{eq:metric-near-P-final}
ds^2\,\Big|_{\text{near-}P}
\sim
- d\widetilde{t}^2 +d\widetilde{y}^2
+ d\widetilde{z}^2 +d\widetilde{x}^2 \,.
\eeq
which corresponds to a patch of Minkowski spacetime.

Observe that the coordinate transformation
\eqref{eq:new-coord-final-y-tilde}
is a $C^1$ function with a discontinuous
second derivative at $y=0$.
That is, the coordinate transformation  is not a
diffeomorphism, which is a $C^\infty$ function everywhere.
The standard elementary-flatness condition
relies, however, on genuine diffeomorphisms.
Hence, the metric \eqref{eq:new-solution-region-III}
obeys a weakened version of the elementary-flatness condition,
allowing for non-smooth coordinate transformations
[the same conclusion holds for the metrics
\eqref{eq:new-BH-solution-PVcoord-chart-123}   
from Painlev\'{e}--Gullstrand-type coordinates].
In other words, the spacetime does not correspond
to a Lorentzian manifold.\footnote{The
solutions of the Klein--Gordon equation over the nonsingular defect manifold
with the \mbox{$M=Q=0$} metric \eqref{eq:new-solution-region-III}
differ from those of Minkowski spacetime~\cite{KlinkhamerSorba2014}.}
Whether or not such spacetimes
play a role in classical physics may ultimately be up to
quantum gravity to decide.



\end{document}